\documentclass[10pt,superscriptaddress,twocolumn,amssymb,amsmath,aps,prl,nofootinbib,notitlepage,floatfix,longbibliography]{revtex4-2}

\usepackage{verbatim}
\usepackage[T1]{fontenc}
\usepackage[utf8]{inputenc}
\usepackage{graphicx}
\usepackage{booktabs}
\usepackage[normalem]{ulem}

\usepackage{bm}
\usepackage{xcolor}
\usepackage{hyperref}

\definecolor{royalblue}{rgb}{0.25,0.41,0.88}
\hypersetup{colorlinks=true,linkcolor=royalblue,citecolor=magenta,urlcolor=royalblue}

\newcommand{\Mpl}{M_{\rm Pl}}

\newcommand{\dd}{\mathrm{d}}

\newcommand{\artanh}{\operatorname{artanh}}

\begin{document}

\title{Inverse Reconstruction of Two-Field Potentials for a DESI-Motivated Phantom-Divide Transition}

\author{Tonghua Liu}
\affiliation{School of Physics and Optoelectronic Engineering, Yangtze University, Jingzhou 434023, China}

\author{Xiaolei Li}
\email{lixiaolei@hebtu.edu.cn}
\affiliation{College of Physics, Hebei Normal University, Shijiazhuang 050024, China}

\author{Marek Biesiada}
\affiliation{National Centre for Nuclear Research, Pasteura 7, PL-02-093 Warsaw, Poland}

\author{Tian-Nuo Li}
\email{litiannuo@ucas.ac.cn}
\affiliation{School of Fundamental Physics and Mathematical Sciences, Hangzhou Institute for Advanced Study, UCAS, Hangzhou 310024, China}

\author{Jun-Qian Jiang}
\email{jiang@kasi.re.kr}
\affiliation{Korea Astronomy and Space Science Institute, Daejeon 34055, Korea}

\author{Arman Shafieloo}
\email{shafieloo@kasi.re.kr}
\affiliation{Korea Astronomy and Space Science Institute, Daejeon 34055, Korea}
\affiliation{University of Science and Technology, 217 Gajeong-ro, Yuseong-gu, Daejeon 34113, Korea}

\author{Xin Zhang}
\email{zhangxin@neu.edu.cn}
\affiliation{Liaoning Key Laboratory of Cosmology and Astrophysics, College of Sciences, Northeastern University, Shenyang 110819, China}
\affiliation{MOE Key Laboratory of Data Analytics and Optimization for Smart Industry, Northeastern University, Shenyang 110819, China}
\affiliation{National Frontiers Science Center for Industrial Intelligence and Systems Optimization, Northeastern University, Shenyang 110819, China}

\date{\today}

\begin{abstract}
A phantom-divide transition favored by recent cosmological observations poses an inverse problem: which field potentials realize the inferred expansion history, and what features are uniquely fixed by it? We reconstruct local potentials for a canonical–phantom scalar pair with standard matter and radiation, targeting Metastable Emergent Dark Energy histories constrained by CMB, DESI BAO, and Type Ia supernova data. The coupled-field Friedmann evolution reproduces the target background to numerical accuracy across the transition, with regular perturbations through the crossing. At the phantom divide, the transition slope sets a dimensionless potential-energy release rate insensitive to the partition of kinetic energy. We identify local potential deformations that preserve both background expansion and linear dynamics. This construction gives a classical field realization of the inferred transition and disentangles observationally fixed energy balances from interaction terms requiring additional theoretical input.
\end{abstract}

\maketitle

\textit{Motivation.---}
Recent Dark Energy Spectroscopic Instrument (DESI) baryon acoustic oscillations (BAO) measurements, combined with cosmic microwave background (CMB) and Type Ia supernova (SN) data, favor evolving dark energy in several analyses~\cite{2025JCAP...02..021A,2025PhRvD.112h3515A,2025NatAs...9.1879G,2025PhRvD.112h3511L}.
The statistical significance depends on the SN sample and the assumed parametrization of its cosmic history.
For noninteracting dark energy, a common feature is a maximum in $\rho_{\rm de}$ as the Universe expands, corresponding to a transition in the dark-energy equation of state from $w<-1$ in the past to $w>-1$ today, which is known as phantom-divide crossing.


A single minimally coupled canonical scalar field cannot accommodate this phantom-divide crossing. Two-field quintom models offer a classical realization using one canonical and one negative-kinetic (phantom) scalar~\cite{2005PhLB..607...35F,2005CoTPh..44..762Z,2006PhLB..634..101F,2010PhR...493....1C}. Alternative interpretations include dark-sector energy transfer~\cite{2024PhRvL.133y1003G,2025JCAP...12..018L,2026arXiv260107361L}, nonminimal gravitational couplings~\cite{2025PhRvL.134r1002Y,2026PhRvD.113d1301T}, derivative self-interactions, or modified gravity~\cite{2005PhRvD..71b3515V,2026PhRvD.113l2006C,2024PhRvD.110l3524C}. Potential reconstruction and the regularity of quintom perturbations have been investigated for many years~\cite{2006PhRvD..73l3509Z,2005PhRvD..72l3515Z}. Two-field inverse reconstruction is known to admit multiple distinct field trajectories and potentials compatible with identical background expansion~\cite{2008JCAP...02..015A}. Recent forward-modeling studies adopt a fixed potential form to realize and constrain phantom-to-nonphantom transitions~\cite{2025MNRAS.544.3142G,2026MNRAS.551g1403G}. By contrast, the inverse approach pursued here begins directly from observationally constrained expansion histories and identifies the potentials consistent with them.

\begin{figure}[!tbp]
\centering
\includegraphics[width=\columnwidth]{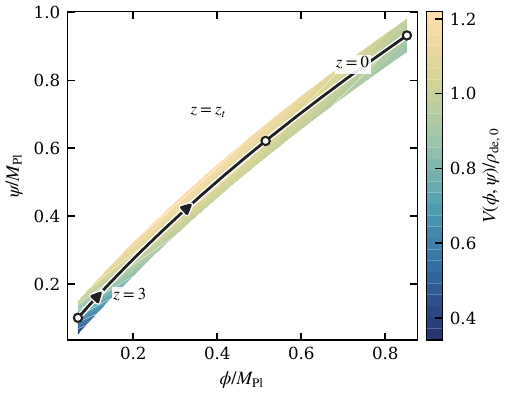}
\caption{One reconstructed potential and its cosmological field trajectory. Shading shows $V(\phi,\psi)/\rho_{{\rm de},0}$ from Eq.~(\ref{eq:potential_main}) near the reference MEDE path, with $\varsigma=0.5$ and $\mu_\perp/H_0=2$. The black curve follows the rolling fields from $z=3$ to $z=0$; arrows indicate increasing cosmic time, and the circle labeled $z=z_t$
marks the phantom-divide crossing. The potential shown is one member of the family defined by Eq.~(\ref{eq:potential_freedom}).}
\label{fig:mede_field_space_potential}
\end{figure}

In this work, we use the observationally constrained Metastable Emergent Dark Energy (MEDE)~\cite{2026ApJ..1001L..21L} expansion histories as targets for inverse reconstruction.
Holding the field-space kinetic metric and standard cosmic species fixed, we construct a local two-field potential and check both the background and perturbations against the observations. Figure~\ref{fig:mede_field_space_potential} displays a representative reconstructed potential together with its cosmological field trajectory. The slope of the inferred equation of state fixes the dimensionless potential-energy release rate, irrespective of how kinetic energy is partitioned between the two fields. This relation maps the inferred transition parameters onto the energy balance for any such field realization. We further isolate families of interaction terms that preserve both the background expansion and perturbation dynamics, separating quantities fixed by the inferred history from choices in the field realization.

\textit{Inverse reconstruction of the two-field potential.---}
We assume a spatially flat Friedmann--Lema\^{i}tre--Robertson--Walker (FLRW) spacetime in general relativity.
Consider two scalar fields $q^I=(\phi,\psi)$ in a field space endowed with the metric $G_{IJ}=\mathrm{diag}(1,-1)$.
Their scalar Lagrangian is
\begingroup
\begin{equation}
\mathcal L=-\frac12G_{IJ}\partial_\mu q^I\partial^\mu q^J-V.
\label{eq:quintom_action_main}
\end{equation}
\endgroup
We use the spacetime signature $(-,+,+,+)$ and units $c=\hbar=1$. The canonical field $\phi$ and phantom scalar field $\psi$ share the potential $V(\phi,\psi)$. For homogeneous fields, $\rho_q=K+V$ and $p_q=K-V$, where $K=(\dot\phi^2-\dot\psi^2)/2$.
The expansion history then determines the dark-energy density $\rho_q$ and pressure $p_q$, and hence
\begingroup
\begin{equation}
\begin{aligned}
V_{\rm tr}=\frac{\rho_q-p_q}{2}=\frac{1-w_q}{2}\rho_q,\quad
K=\frac{\rho_q+p_q}{2}.
\end{aligned}
\label{eq:rho_p_q_main}
\end{equation}
\endgroup
Here $\mathrm{tr}$ labels the cosmological trajectory,  $H=\dot a/a$, $a=1/(1+z)$, $M_{\rm Pl}=(8\pi G)^{-1/2}$,
and an overdot denotes differentiation with respect to cosmic time. For a fixed expansion history $H(z)$, the inferred dark-energy equation of state $w_q=p_q/\rho_q$ depends on the adopted composition of standard matter and radiation.
The expression for $V_{\rm tr}$ applies to any number of minimally coupled scalar fields with quadratic kinetic terms.

The homogeneous evolution traces a trajectory $\gamma:t\mapsto(\phi(t),\psi(t))$ in this field space. The background determines $K(t)$ and $V_{\rm tr}(t)$, while the individual field velocities remain free.
To choose a trajectory, specify their positive kinetic sum $S_K=(\dot\phi^2+\dot\psi^2)/2>|K|$ and set
\begin{equation}
\dot\phi=\sqrt{S_K+K},\qquad \dot\psi=\sqrt{S_K-K}.
\label{eq:split_main}
\end{equation}
We take $S_K$ smooth and use positive velocities; other sign choices are possible.  Integrating Eq.~(\ref{eq:split_main}) gives a monotonic path $\psi=\psi_{\rm tr}(\phi)$.  At the phantom-divide crossing, the trajectory tangent is null with respect to the indefinite field-space metric  $G_{IJ}$ even though both fields remain in motion; this null condition corresponds to equal magnitudes of the two kinetic terms. Different choices of $S_K$ give different field trajectories with the same expansion.

The inferred $V_{\rm tr}(t)$ gives the potential along this trajectory.  A field theory also needs its gradients and its behavior in neighboring field configurations.  Define $U(\phi)=V_{\rm tr}[t(\phi)]$ and $J(\phi)=[\ddot\psi+3H\dot\psi]_{t=t(\phi)}$.  One local potential with the required value and gradients is
\begin{equation}
V(\phi,\psi)=U(\phi)+J(\phi)y-\frac12\mu_\perp^2y^2,
\quad y=\psi-\psi_{\rm tr}(\phi).
\label{eq:potential_main}
\end{equation}
Here  $J(\phi)$ gives the potential gradient along $\psi$ at fixed $\phi$. The scale $\mu_\perp$ sets the transverse curvature $V_{,\psi\psi}=-\mu_\perp^2$ evaluated on the trajectory, but it is not a physical mass eigenvalue of the coupled two-field perturbation system. On the trajectory, $V=U$ and $V_{,\psi}=J$, enforcing the phantom Klein--Gordon equation.  The canonical equation follows from energy conservation and the relation $V_{,\phi}=U_{,\phi}-J\psi_{{\rm tr},\phi}$.  Thus the inferred history is a solution of
\begin{equation}
\begin{aligned}
&\ddot q^I+3H\dot q^I+G^{IJ}V_{,J}=0,\\
&3M_{\rm Pl}^2H^2=\rho_{\rm std}+K+V ,
\end{aligned}
\label{eq:forward_main}
\end{equation}
where $G^{IJ}$ is the inverse field-space metric, a comma denotes a field derivative, and repeated indices are summed over.  Once $U$, $J$, and $\psi_{\rm tr}$ are fixed, Eq.~(\ref{eq:potential_main}) depends only on the field positions.  

For this fixed trajectory, the remaining local freedom in the potential has the form
\begin{equation}
\widetilde V=V+\frac12B(\phi)y^2+y^3C(\phi,y),
\label{eq:potential_freedom}
\end{equation}
where $B(\phi)$ and $C(\phi,y)$ are smooth functions with mass dimensions $2$ and $1$, respectively.
The added potential and its first gradient vanish identically at $y=0$, so the background solution remains unaltered. A nonzero quadratic term changes the curvature of the potential across the trajectory and can alter linear perturbations. If $B(\phi)=0$, the two potentials also have the same second derivatives along the trajectory. For identical initial perturbations, the fields, metric, and standard matter components then evolve identically at linear order, although the forces can differ away from the trajectory. Thus even the background and linear
perturbations together do not determine the potential away from the trajectory. A similar freedom beyond linear order arises in Horndeski reconstruction~\cite{2017PhRvD..96h4051K,2019PhRvD.100d4034K}. The Supplemental Material (SM)~\cite{SupplementalMaterial}, Sec.~\ref{app:quintom}, gives the local derivation using Hadamard's lemma and illustrates the cubic case in Fig.~\ref{fig:equivalent_potentials}.

\textit{Reconstruction of the MEDE history.---}
We use the natural-logarithm MEDE parametrization for the target expansion history:
\begin{equation}
w_{\rm de}(z)=-1-\Delta\tanh u(z),\qquad
u(z)=\ln\frac{1+z}{1+z_t},
\label{eq:wz_main}
\end{equation}
whose density follows from energy conservation:
\begin{equation}
\rho_{\rm de}(z)=\rho_{{\rm de},0}
\left[\frac{\cosh u(z)}{\cosh u(0)}\right]^{-3\Delta}.
\label{eq:rho_main}
\end{equation}
The transition redshift $z_t$ locates the density maximum for variation amplitude $\Delta>0$. SM Sec.~\ref{app:background} relates this convention to the original base-10 form and discusses the $\Delta=0$ limit.
\begin{figure}[!tbp]
\centering
\includegraphics[width=\columnwidth]{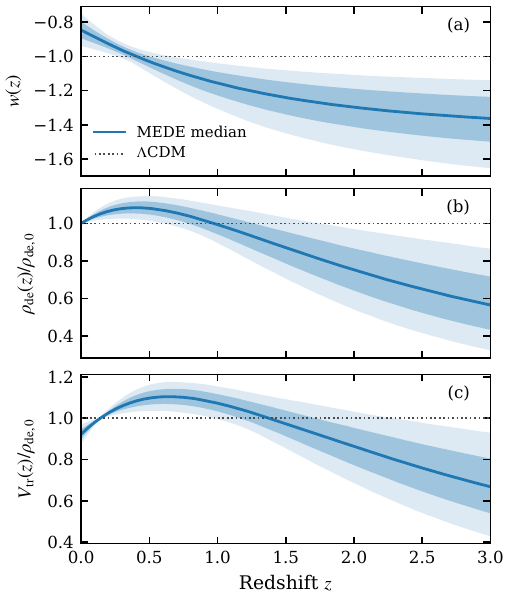}
\caption{MEDE histories and the reconstructed potential along the field path from CMB+DESI DR2 BAO+DES-Dovekie.  Panels show (a) $w(z)$, (b) $\rho_{\rm de}(z)$, and (c) $V_{\rm tr}(z)=[1-w(z)]\rho_{\rm de}(z)/2$, with energies normalized by $\rho_{{\rm de},0}$.  Curves and bands give the median and pointwise $68\%$ and $95\%$ credible intervals from the weighted chains after burn-in.  Dotted lines indicate the $\Lambda$CDM values.  Reconstructing $V_{\rm tr}(z)$ requires no choice of the individual field velocities.}
\label{fig:mede_history}
\end{figure}

Our joint Markov chain Monte Carlo (MCMC) analysis combines CMB temperature, polarization, and lensing data, DESI DR2 BAO, and the DES-Dovekie supernova compilation~\cite{2020A&A...641A...6P,2022MNRAS.517.4620R,2022JCAP...09..039C,2024ApJ...962..112Q,2025PhRvD.112h3515A,2026MNRAS.548ag632P}. The chains sample the phenomenological MEDE expansion history.
The DES-Dovekie combination gives $z_t=0.399^{+0.079}_{-0.071}$ and $\Delta=0.492^{+0.150}_{-0.141}$ (see SM Sec.~\ref{app:full_table} for priors, full constraints, and chain diagnostics). Each posterior sample gives $w(z)$ and $\rho_{\rm de}(z)$, from which Eq.~(\ref{eq:rho_p_q_main}) determines $V_{\rm tr}(z)$. Figure~\ref{fig:mede_history} shows the resulting histories and their credible bands.

\begin{figure}[!tbp]
\centering
\includegraphics[width=\columnwidth]{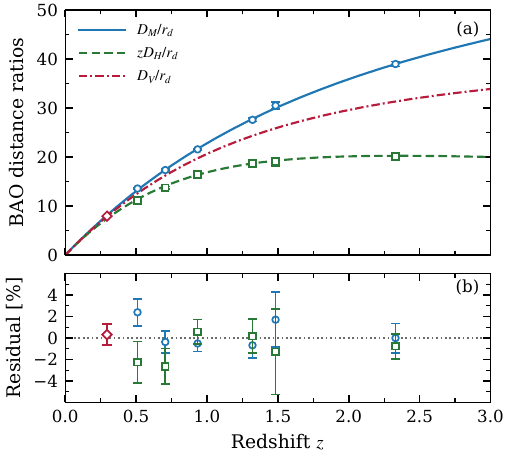}
\caption{(a) The solid, dashed, and dash-dotted curves give the MEDE posterior medians of $D_M/r_d$, $zD_H/r_d$, and $D_V/r_d$, respectively, from the same joint posterior as Fig.~\ref{fig:mede_history}. Circles, squares, and diamonds mark the corresponding DESI DR2 BAO measurements~\cite{2025PhRvD.112h3515A}, with marginal $1\sigma$ errors from the Gaussian likelihood covariance. For display, both the radial distance ratio $D_H/r_d$ and its error are multiplied by $z$. (b) Percentage residuals $100(X_{\rm DESI}/X_{\rm MEDE}-1)$, with $X_{\rm MEDE}$ the median of the corresponding ratio at the measurement redshift. The error bars are the observational errors expressed on the same percentage scale.} 
\label{fig:bao_distances}
\end{figure}

The same posterior samples give the MEDE distances in Fig.~\ref{fig:bao_distances}. The lower panel shows the residuals of the DESI DR2 measurements used in the fit relative to the MEDE medians. In a flat universe, the radial distance $D_H=c/H(z)$ measures the expansion at $z$, while the transverse comoving distance $D_M=\int_0^z D_H(z')\,\mathrm{d}z'$ depends on the expansion up to that redshift. The lowest-redshift bin measures the angle-averaged distance $D_V=(zD_M^2D_H)^{1/3}$. The three distance ratios use the sound horizon $r_d$ at baryon drag as a common ruler.

For an explicit field realization, we use the joint-posterior median parameter set $z_t=0.399$, $\Delta=0.492$, and $\Omega_{\rm m0}=0.3121$. This is a representative field realization of the posterior. We take $S_K=\sqrt{K^2+\varsigma^2\rho_{\rm de}^2}$ with $\varsigma=0.5$ and $\mu_\perp/H_0=2$. These choices specify the individual field motion and the curvature away from the path. Figure~\ref{fig:mede_field_space_potential} shows the resulting potential from Eq.~(\ref{eq:potential_main}), and SM Fig.~\ref{fig:forward_dynamics} shows its functions $U(\phi)$ and $J(\phi)$.

With this potential and the corresponding initial field state fixed, we solve the coupled Klein--Gordon--Friedmann system in Eq.~(\ref{eq:forward_main}) with the modified Cosmic Linear Anisotropy Solving System (CLASS)~\cite{2011JCAP...07..034B}. The expansion follows from the rolling fields and the fixed standard species. The forward solution reproduces the target MEDE history. SM Sec.~\ref{app:forward} gives the initial data and early-time continuation.

The field motion explains the density turnover.  Before the crossing, $\dot\psi^2>\dot\phi^2$, so $K<0$ and energy conservation gives $\dot\rho_q=-6HK>0$: the dark-energy density grows as the Universe expands.  At $z=z_t$, the two kinetic magnitudes are equal.  The canonical contribution then becomes larger, $K>0$, and the density decreases.  Both fields continue to roll throughout.  Their motion is coupled by the gradients of the shared potential, so the transition is a change in kinetic balance rather than the disappearance of the phantom field.

The potential is already decreasing when the field density reaches its maximum. At the crossing, $\dot\rho_q=0$ and $K$ is increasing, so $\dot V_{\rm tr}=-\dot K<0$. For the representative solution, the potential peaks at $z_{\max}=0.638$, before the crossing at $z_t=0.399$, as shown in SM Figs.~\ref{fig:forward_dynamics}(a) and~\ref{fig:quintom_reconstruction}(b). The dimensionless release rate $\mathcal R_t=-\dot V_{\rm tr}/(H\rho_q)|_{z_t}$ is half the slope of $w_q$ with respect to $\ln a$ at the crossing [SM Eq.~(\ref{eq:turnover_work})]. For the MEDE history in Eq.~(\ref{eq:wz_main}), this gives $\mathcal R_t=\Delta/2$, with $0.246^{+0.075}_{-0.071}$ at $68\%$ credibility for the joint combination (SM Table~\ref{tab:full_results}).

To predict the growth of structure, we evolve both field fluctuations together with matter, radiation, and metric perturbations. The fields couple through the potential Hessian $V_{,IJ}$ and the metric, and their equations remain finite at the phantom-divide crossing $w_q=-1$. We use adiabatic initial conditions for the standard species and set the independent initial field fluctuations and their derivatives to zero (SM Sec.~\ref{app:perturbations}).

\begin{figure}[!tbp]
\centering
\includegraphics[width=\columnwidth]{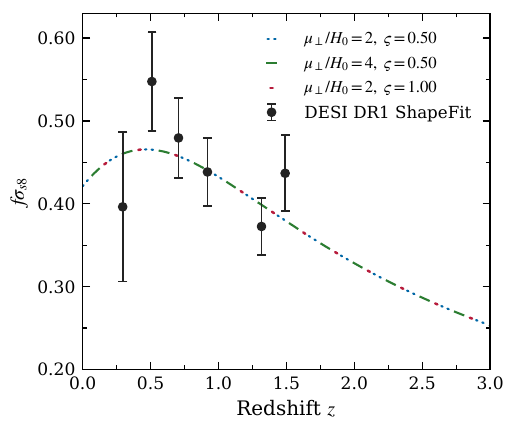}
\caption{DESI DR1 ShapeFit measurements of the growth quantity $f\sigma_{s8}$ (black circles with marginal $1\sigma$ errors) and predictions for three two-field realizations of the same MEDE expansion history. The reference has $(\mu_\perp/H_0,\varsigma)=(2,0.50)$ (blue paired dots). Here $\mu_\perp/H_0$ controls the potential curvature away from the trajectory at fixed $\phi$, and $\varsigma$ sets the kinetic sum $S_K$.
The other two cases vary the curvature, $(4,0.50)$ (green dashes), or the kinetic sum, $(2,1.00)$ (red dots). At the measured redshifts, the differences among the three predictions are much smaller than the observational errors. The calculation is described in SM Sec.~\ref{app:shapefit}.}
\label{fig:field_linear_shapefit}
\end{figure}

As an example, Figure~\ref{fig:field_linear_shapefit} compares estimates of the compressed growth quantity $f\sigma_{s8}$ [SM Eq.~(\ref{eq:shapefit_growth})] for three realizations of the same MEDE expansion with DESI DR1 ShapeFit measurements~\cite{2025JCAP...09..008A}. Relative to the reference, one realization changes the curvature away from the trajectory by increasing $\mu_\perp/H_0$ from $2$ to $4$ at fixed $\varsigma$. The other changes the kinetic sum $S_K$ by increasing $\varsigma$ from $0.50$ to $1.00$
at fixed $\mu_\perp/H_0$, changing both the trajectory and its potential. The differences among the three predictions are much smaller than the measurement errors (SM Sec.~\ref{app:shapefit}). This comparison does
not enter the likelihood. The near overlap in
Fig.~\ref{fig:field_linear_shapefit} is distinct from the exact linear equivalence for cubic potential deformations in
Eq.~(\ref{eq:potential_freedom}).

\textit{Conclusion.---}
We have reconstructed a local two-field potential that realizes a data-constrained MEDE history through the coupled field and Friedmann equations, with regular linear perturbations at the crossing.  Both fields remain in motion as the canonical kinetic energy overtakes the phantom kinetic magnitude and the density begins to decrease.  The history fixes $V_{\rm tr}(t)$ and the instantaneous potential-energy release rate at the crossing, independently of the chosen kinetic split.  The MEDE posterior therefore constrains this net energy balance without determining the individual field motions or the complete interaction.  For a given trajectory and fixed kinetic terms, cubic potential deformations can change the forces away from the path while preserving the same background and linear perturbations for identical initial conditions. The pathologies associated with the negative kinetic term and the limited validity of this purely classical description are discussed in SM Sec.~\ref{app:scope}.

\textit{Acknowledgments.---}
We thank Ryan E. Keeley for helpful discussions and valuable suggestions. X.L. is supported by the Natural Science Foundation of Hebei Province under Grant No.~A2026205012 and the National Natural Science Foundation of China (No. 12673002). X.Z. is supported by the National Natural Science Foundation of China (Grants Nos. 12533001 and 12473001), the National SKA Program of China (Grants Nos. 2022SKA0110200 and 2022SKA0110203), and the 111 Project (Grant No. B16009).
J.J. and A.S. acknowledge funding from the Korea Astronomy and Space Science Institute (KASI) through the project ``Research on the Principles of the Accelerating Expansion of the Universe'' (Project Code: 2026183201).
A.S. acknowledges support from the Korea Institute for Advanced Study grant funded by the Korean government.  M.B. was supported by the Polish National Science Centre grant 2023/50/A/ST9/00579.

\textit{Data availability.---}
{The observational data used here are publicly available from the repositories cited in the references. Posterior chains, CLASS configuration files, field lookup tables, and numerical scripts used to generate our results are available from the corresponding authors upon reasonable request.}

\bibliography{ref}

\clearpage
\onecolumngrid

\begin{center}
\huge{\textsc{Supplemental Material}}

\vspace{0.6em}
{\large\textbf{Inverse Reconstruction of Two-Field Potentials for a DESI-Motivated Phantom-Divide Transition}}

\vspace{0.4em}
{\normalsize Tonghua Liu, Xiaolei Li, Marek Biesiada, Tian-Nuo Li, Jun-Qian Jiang, Arman Shafieloo, and Xin Zhang}
\end{center}



\newcounter{supsection}
\renewcommand{\thesupsection}{S\arabic{supsection}}

\newcommand{\suppsection}[1]{%
  \refstepcounter{supsection}%
  \section*{\thesupsection.\ #1}%
}

\newcounter{supsubsection}[supsection]
\renewcommand{\thesupsubsection}{\thesupsection.\arabic{supsubsection}}

\newcommand{\suppsubsection}[1]{%
  \refstepcounter{supsubsection}%
  \subsection*{\thesupsubsection.\ #1}%
}

\setcounter{figure}{0}
\renewcommand{\thefigure}{S\arabic{figure}}
\renewcommand{\theHfigure}{S.\arabic{figure}}
\setcounter{table}{0}
\renewcommand{\thetable}{S\arabic{table}}
\renewcommand{\theHtable}{S.\arabic{table}}
\setcounter{equation}{0}
\renewcommand{\theequation}{S\arabic{equation}}
\renewcommand{\theHequation}{S.\arabic{equation}}

\suppsection{Background evolution}
\label{app:background}

For fixed standard species, the Friedmann equations determine the dark-energy density and pressure from the expansion history:
\begin{align}
&\rho_q=3M_{\rm Pl}^2H^2-\rho_{\rm std},\nonumber\\
&p_q=-M_{\rm Pl}^2(2\dot H+3H^2)-p_{\rm std}.
\label{eq:history_inference}
\end{align}

The dimensionless expansion rate associated with Eqs.~(\ref{eq:wz_main}) and (\ref{eq:rho_main}) is
\begin{equation}
E^2(z)=E_{\rm std}^2(z)+\Omega_{{\rm de},0}f_{\rm de}(z),
\quad E=H/H_0,\quad
E_{\rm std}^2=\frac{\rho_{\rm std}}{3M_{\rm Pl}^2H_0^2},
\quad f_{\rm de}=\frac{\rho_{\rm de}}{\rho_{{\rm de},0}}.
\label{eq:Ez_main}
\end{equation}
Here $\Omega_{{\rm de},0}=\rho_{{\rm de},0}/(3M_{\rm Pl}^2H_0^2)$ and the standard matter species are those used in the joint likelihood.  Matter and dark energy densities are separately conserved.

For $\Delta>0$, MEDE is phantom-like before the transition and nonphantom afterwards.  We derive the density evolution and its asymptotic limits.

Starting from Eq.~(\ref{eq:wz_main}), the dark-energy continuity equation is
\begin{equation}
\dot\rho_{\rm de}+3H(1+w_{\rm de})\rho_{\rm de}=0.
\label{eq:de_continuity}
\end{equation}
Using $\dot z=-H(1+z)$,
\begin{equation}
\frac{\dd\ln\rho_{\rm de}}{\dd z}=\frac{3[1+w_{\rm de}(z)]}{1+z}=-\frac{3\Delta}{1+z}\tanh u(z).
\label{eq:de_continuity_redshift}
\end{equation}
Since $\dd u/\dd z=(1+z)^{-1}$,
\begin{equation}
\frac{\dd\ln\rho_{\rm de}}{\dd u}=-3\Delta\tanh u,
\label{eq:de_continuity_u}
\end{equation}
which integrates to Eq.~(\ref{eq:rho_main}).  The limit $\Delta=0$ is exactly $\Lambda$CDM; the density is constant and independent of $z_t$, so its transition redshift cannot be constrained since it is not even properly defined.

The asymptotic limits are
\begin{align}
&z\gg z_t: \quad w_{\rm de}\to -1-\Delta,\qquad \rho_{\rm de}\propto(1+z)^{-3\Delta},
\label{eq:early_time_asymptote}
\\
&a\to\infty: \quad w_{\rm de}\to -1+\Delta,\qquad \rho_{\rm de}\propto a^{-3\Delta}.
\label{eq:future_asymptote}
\end{align}
Relative to matter, $\rho_{\rm de}/\rho_m\propto(1+z)^{-3(1+\Delta)}$ at high redshifts.

{More generally,} a width parameter $\beta>0$ can be introduced as
\begin{equation}
w_{\rm de}(z)=-1-\Delta\tanh\left[\beta\ln\left(\frac{1+z}{1+z_t}\right)\right],
\label{eq:variable_width_eos}
\end{equation}
which gives
\begin{equation}
\rho_{\rm de}(z)=\rho_{{\rm de},0}\left[\frac{\cosh[\beta\,u(z)]}{\cosh[\beta\,u_{0}]}\right]^{-3\Delta/\beta}.
\label{eq:variable_width_density}
\end{equation}
Here $u_0=u(0)$.  The main analysis fixes $\beta=1$ and does not fit the transition width independently.

 {Equation~(\ref{eq:wz_main}) uses a natural logarithm, whereas the original MEDE analysis used a decimal logarithm~\cite{2026ApJ..1001L..21L}.  {For fixed $(z_t,\Delta)$, changing the logarithm base rescales the argument of the hyperbolic tangent and hence the transition width, while $\Delta$ remains the variation amplitude.}  The constraints and field reconstruction reported here therefore refer to the natural-logarithm convention defined in Eq.~(\ref{eq:wz_main}).}

\suppsection{Priors and full likelihood table}
\label{app:full_table}

\begin{figure}[htbp]
\centering
\includegraphics[width=0.48\textwidth]{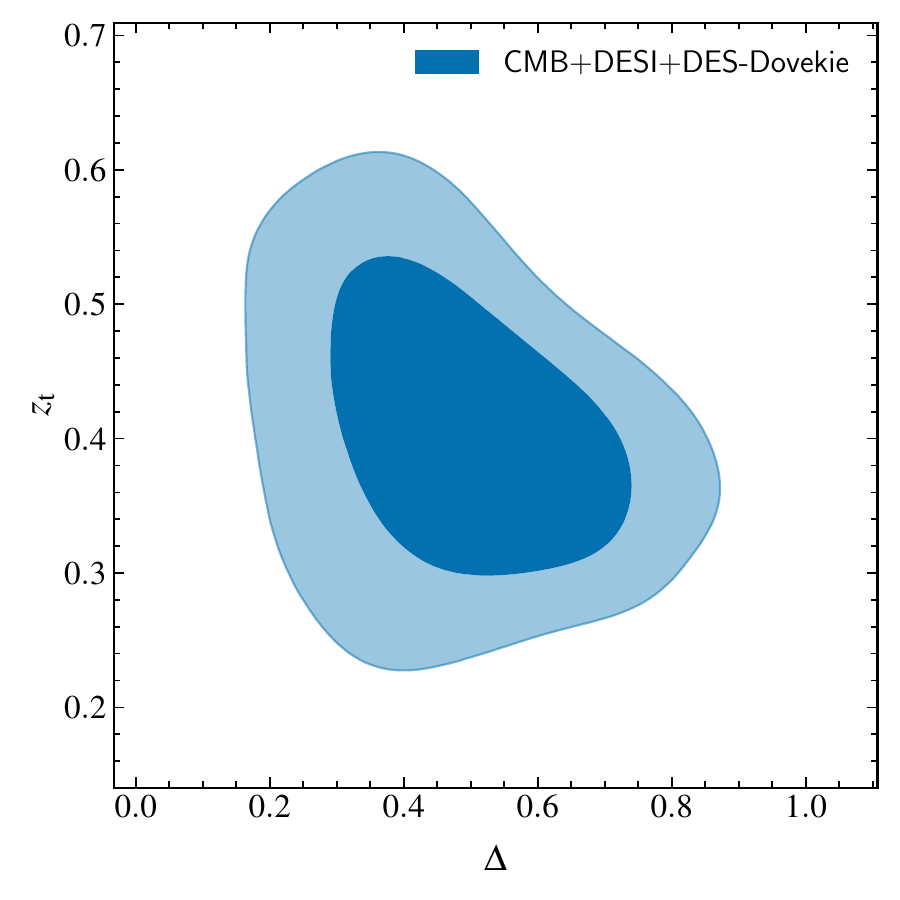}
\caption{Marginalized MEDE constraints on the transition redshift $z_t$ and variation amplitude $\Delta$ for CMB+DESI+DES-Dovekie.  Inner and outer contours enclose $68\%$ and $95\%$ posterior probability, respectively.}
\label{fig:constraints}
\end{figure}

The chains employ the parametrized post-Friedmann (PPF) framework to describe dark-energy perturbations~\cite{2008PhRvD..78h7303F}. We combine CMB observations from Planck, ACT, and SPT-3G with DESI DR2 BAO and the DES-Dovekie supernova. For the CMB, we adopt the Planck low-$\ell$ TT and EE likelihoods~\cite{Planck:2019nip}, the Planck PR4 NPIPE \texttt{CamSpec} high-$\ell$ TT, TE, and EE likelihood~\cite{2021OJAp....4E...8E,Planck:2020olo,Jense:2025wyg}, the ACT DR6 \texttt{ACT-lite} likelihood~\cite{ACT:2025fju}, and the SPT-3G D1 \texttt{SPT-lite} likelihood~\cite{SPT-3G:2025bzu}. Appropriate multipole cuts are applied to avoid double counting when combining the high-$\ell$ likelihoods; see Ref.~\cite{2026arXiv260727410D} for details. We further incorporate the unified CMB lensing likelihood constructed from the lensing reconstructions of Planck, ACT, and SPT-3G~\cite{ACT:2025qjh}. Figure~\ref{fig:constraints} shows the transition-parameter constraints.  Table~\ref{tab:priors} lists the parameter priors, and Table~\ref{tab:full_results} gives the posterior constraints.

\begin{center}
\begin{minipage}{\textwidth}
\refstepcounter{table}\label{tab:priors}
\footnotesize
\textbf{Table~\thetable.} Priors for the reported MEDE posterior constraints.   {We define $\omega_b\equiv\Omega_{\rm b0}h^2$ and $\omega_{\rm cdm}\equiv\Omega_{\rm cdm0}h^2$}; $\mathcal U[a,b]$ denotes a uniform prior on $[a,b]$.\\[3pt]
\setlength{\tabcolsep}{9pt}
\renewcommand{\arraystretch}{1.08}
\begin{ruledtabular}
\begin{tabular}{llll}
Parameter & Prior & Parameter & Prior \\
\hline
 {$\omega_b$} & $\mathcal U[0.005,0.1]$ & $\tau_{\rm reio}$ & $\mathcal U[0.01,0.8]$ \\
 {$\omega_{\rm cdm}$} & $\mathcal U[{0.001},0.99]$ &  {$\ln(10^{10}A_s)$} & $\mathcal U[1.61,3.91]$ \\
$H_0$ & $\mathcal U[20,100]$ & $n_s$ & $\mathcal U[0.8,1.2]$ \\
$z_t$ & $\mathcal U[0,1]$ & $\Delta$ & $\mathcal U[-1,2]$ \\
\end{tabular}
\end{ruledtabular}
\end{minipage}
\end{center}

\begin{table}[htbp]
\centering
\caption{Posterior medians and central $68\%$ credible intervals for the MEDE model using CMB+DESI+DES-Dovekie. Here $\omega_b$ and $\omega_{\rm cdm}$ are the physical baryon and cold-dark-matter densities; $A_s$ and $n_s$ are the primordial scalar amplitude and tilt; $\tau_{\rm reio}$ is the optical depth to reionization; $\sigma_8$ is the rms matter fluctuation in $8h^{-1}{\rm Mpc}$ spheres; and $r_{\rm drag}$ is the sound horizon at the baryon-drag epoch. We define $\Delta\chi^2_{{\rm MEDE}-j}\equiv\chi^2_{\min,{\rm MEDE}}-\chi^2_{\min,j}$ and use minimum $\chi^2$ values only to compare fit quality. The derived quantity $\mathcal R_t=-\dot V/(H\rho_q)|_{z_t}$ satisfies $\mathcal R_t=\Delta/2$ for this parameterization; its $68\%$ and $95\%$ credible intervals are reported in the table.} 
\label{tab:full_results}
\setlength{\tabcolsep}{5mm}
\renewcommand{\arraystretch}{1.5}
\footnotesize
\resizebox{0.60\columnwidth}{!}{%
\begin{tabular}{lc}
\hline
\hline
\textbf{Parameter} &
\textbf{CMB+DESI+DES-Dovekie} \\
\hline
$z_t$ & $0.399^{+0.079}_{-0.071}$ \\
$\Delta$ & $0.492^{+0.150}_{-0.141}$ \\
\hline
$\omega_b$ & $0.02242^{+0.00009}_{-0.00009}$ \\
$\omega_{\rm cdm}$ & $0.11919^{+0.00072}_{-0.00078}$ \\
$\ln(10^{10}A_s)$ & $3.051^{+0.012}_{-0.012}$ \\
$n_s$ & $0.9715^{+0.0029}_{-0.0029}$ \\
$\tau_{\rm reio}$ & $0.0582^{+0.0066}_{-0.0065}$ \\
\hline
$H_0\,[{\rm km\,s^{-1}\,Mpc^{-1}}]$ &
$67.51^{+0.57}_{-0.50}$ \\
$\Omega_{\rm m0}$ & $0.3121^{+0.0048}_{-0.0052}$ \\
$\sigma_8$ & $0.8161^{+0.0074}_{-0.0071}$ \\
$r_{\rm drag}\,[{\rm Mpc}]$ &
$147.27^{+0.18}_{-0.19}$ \\
\hline
$\chi^2_{\min,{\rm MEDE}}$ & $5688.23$ \\
$\Delta\chi^2_{{\rm MEDE}-\Lambda{\rm CDM}}$ &
$-15.11$ \\
\hline
$\mathcal R_t\,(68\%)$ &
$0.246^{+0.075}_{-0.071}$ \\
$\mathcal R_t\,(95\%\ {\rm interval})$ &
$[0.112,\,0.383]$ \\
\hline
\hline
\end{tabular}%
}
\end{table}

The joint MEDE MCMC analysis used Cobaya{~\cite{2021JCAP...05..057T}} with a custom Code for Anisotropies in the Microwave Background (CAMB){~\cite{2000ApJ...538..473L}} implementation, one massive neutrino with $\sum m_\nu=0.06\,{\rm eV}$, $N_{\rm eff}=3.044$, a spatially flat background, and BBN-consistent helium abundance. The retained chain weights are used after discarding the first $30\%$ of the chain. The field reconstruction focuses on the $\Delta>0$ branch. The reported $\chi^2$ values are best-fit likelihood values from the attached MEDE and $\Lambda$CDM runs; the checkpoint for the MEDE chains satisfy $R-1<0.02$ before completion, so no stronger convergence claim is made here.  We obtain the best-fit $\chi^2=-2\ln\mathcal L$ in a separate BOBYQA optimization in Cobaya, without including prior density in the objective.

\paragraph{Posterior histories and distances.}
Figures~\ref{fig:mede_history} and~\ref{fig:bao_distances} use the retained samples of the DES-Dovekie chains, with the same importance weights as Table~\ref{tab:full_results} and no cut on $\Delta$.  At each redshift, we compute weighted quantiles for
Fig.~\ref{fig:mede_history} and weighted medians for
Fig.~\ref{fig:bao_distances}. Only Fig.~\ref{fig:mede_history}
shows posterior credible bands.  For Fig.~\ref{fig:bao_distances}, we compute the distances from each sample's MEDE expansion and divide them by that sample's $r_d$.  The displayed errors come from the diagonal of the DESI DR2 Gaussian likelihood covariance; the likelihood uses the full covariance.

\paragraph{Potential-energy release at the crossing.}
For each retained row we evaluate $\mathcal R_t=\Delta/2$, using the same $30\%$ burn-in removal and posterior weights as Table~\ref{tab:full_results}.  We retain both signs of $\Delta$: positive $\mathcal R_t$ describes release at a density maximum, whereas negative $\mathcal R_t$ describes absorption at a minimum.  The intervals in Table~\ref{tab:full_results} use the weighted empirical distribution, with no extra likelihood or branch cut.  With the variable-width history in Eq.~(\ref{eq:variable_width_eos}), the relation would be $\mathcal R_t=\beta\Delta/2$.  Our numerical constraints are conditional on $\beta=1$ and the MEDE likelihood.


\suppsection{Scalar-quintom background reconstruction}
\label{app:quintom}

\begin{figure*}[t]
\centering
\includegraphics[width=0.92\textwidth]{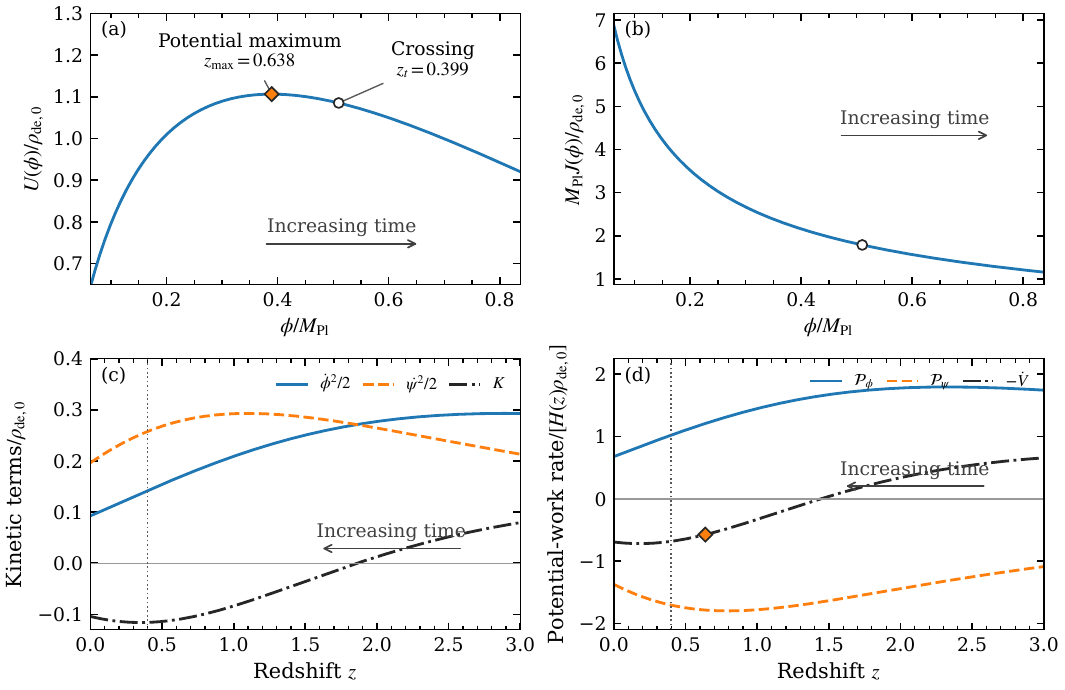}
\caption{Potential and kinetic evolution of the representative solution with $z_t=0.399$, $\Delta=0.492$, and $\Omega_{\rm m0}=0.3121$.  (a) $U(\phi)$ is the potential on the reconstructed path; the diamond is placed at the numerical maximum of $U(\phi)$, corresponding to $z_{\max}=0.638$; (b) $J(\phi)$ is its slope in the $\psi$ direction.  The open circle in (a) marks the later crossing at $z_t=0.399$.  Cosmic time increases toward larger $\phi$ but smaller $z$, as indicated by the arrows.  (c) The positive kinetic magnitudes $\dot\phi^2/2$ and $\dot\psi^2/2$, and their difference $K$.  (d) The potential-work rates $\mathcal P_\phi=-V_{,\phi}\dot\phi$ and $\mathcal P_\psi=-V_{,\psi}\dot\psi$, with sum $-\dot V$; the diamond marks its zero at the potential maximum.  Dotted lines mark the crossing.  Curves use the reconstructed background solution over $0\leq z\leq3$.  Energies are normalized by $\rho_{{\rm de},0}$ and power terms by $H(z)\rho_{{\rm de},0}$.}
\label{fig:forward_dynamics}
\end{figure*}

{We reconstruct a potential from an allowed expansion history and then evolve the fields in that fixed potential.  The fitted MEDE functions $w_{\rm de}(z)$ and $\rho_{\rm de}(z)$ determine the kinetic-energy difference $K(z)$ and the potential $V_{\rm tr}(z)$ along a cosmological trajectory.  Specifying the positive kinetic sum gives the individual velocities and hence the trajectory.  The field equations then fix the potential gradients on this curve.  Extending the potential to its neighborhood defines the forces for displaced fields and the evolution of their perturbations.  The freedom in choosing both the trajectory and this extension is a feature of two-field reconstruction~\cite{2008JCAP...02..015A}.}

 {The scalar-quintom realization is a homogeneous reconstruction of the MEDE trajectory, comprising a canonical field $\phi$ and a phantom scalar field $\psi$ with action}
\begin{equation}
 S=\int \dd^4x\sqrt{-g}\left[\frac{\Mpl^2}{2}R-\frac{1}{2}(\partial\phi)^2+\frac{1}{2}(\partial\psi)^2-V(\phi,\psi)\right]+S_m+S_r,
\label{eq:quintom_action_supp}
\end{equation}
 {where $R$ is the Ricci scalar, and $S_m$ and $S_r$ denote the matter and radiation actions.  {The scalar Lagrangian in Eq.~(\ref{eq:quintom_action_supp}) is given by Eq.~(\ref{eq:quintom_action_main}).}  The opposite kinetic signs allow $\rho_q+p_q=\dot\phi^2-\dot\psi^2$ to pass through zero while both fields continue to roll.}
For homogeneous fields,
\begin{align}
&\rho_{\rm de}=\frac12\dot\phi^2-\frac12\dot\psi^2+V,
\label{eq:field_density_supp}
\\
&p_{\rm de}=\frac12\dot\phi^2-\frac12\dot\psi^2-V.
\label{eq:field_pressure_supp}
\end{align}
We define $\Omega_q=\rho_q/(3M_{\rm Pl}^2H^2)$ as the fractional field density.  Let $K$ denote the kinetic-energy difference between the two fields,
\begin{equation}
K=\frac12\dot\phi^2-\frac12\dot\psi^2=\frac{1+w_{\rm de}}{2}\rho_{\rm de},\qquad V=\frac{1-w_{\rm de}}{2}\rho_{\rm de}.
\label{eq:reconstruction_identity}
\end{equation}
For MEDE,
\begin{align}
&K(z)=-\frac{\Delta}{2}\tanh u(z)\rho_{\rm de}(z),
\label{eq:mede_kinetic_supp}
\\
&V(z)=\left[1+\frac{\Delta}{2}\tanh u(z)\right]\rho_{\rm de}(z).
\label{eq:mede_potential_supp}
\end{align}
The potential maximum and the density maximum occur at different times.  At $z=z_t$, $K=0$ but $\dot K=\Delta H\rho_q/2>0$, so $\dot\rho_q=0$ and $\dot V=-\dot K<0$.  The potential has already passed its maximum.  For $\Delta>0$, differentiating Eq.~(\ref{eq:mede_potential_supp}) gives
\begin{equation}
z_{\max}=(1+z_t)\exp\!\left[\artanh\!\left(\frac{1}{3+\sqrt{10+3\Delta}}\right)\right]-1 .
\label{eq:potential_peak_redshift}
\end{equation}
The representative parameters give $z_{\max}=0.638$, earlier than the crossing at $z_t=0.399$.  Along the reconstructed trajectory, $U[\phi(z)]=V(z)$, so the maxima in Figs.~\ref{fig:forward_dynamics}(a) and \ref{fig:quintom_reconstruction}(b) are related to the same event.  Their horizontal axes run oppositely in time: $\phi$ increases while $z$ decreases.  At the crossing $V=\rho_q$, but equality at that instant does not imply coincident extrema.

Half of the sum of the squared velocities, $S_K=(\dot\phi^2+\dot\psi^2)/2$, is an additional choice.  Choose it smooth with $S_K(z)>|K(z)|$ and set
\begin{equation}
\dot\phi^2=S_K+K,
\qquad
\dot\psi^2=S_K-K.
\label{eq:kinetic_split_supp}
\end{equation}
Then
\begin{equation}
\frac{\dd\phi}{\dd z}=-\frac{\sqrt{S_K+K}}{(1+z)H(z)},
\qquad
\frac{\dd\psi}{\dd z}=-\frac{\sqrt{S_K-K}}{(1+z)H(z)}.
\label{eq:field_path_redshift}
\end{equation}
 {A smooth choice is $S_K=\sqrt{K^2+\varsigma^2\rho_{\rm de}^2}$ with $\varsigma>0$.  At fixed $K$ and $V$, $\varsigma$ changes the individual kinetic energies but leaves their difference, and thus the target density and pressure, unchanged.  To define the forces when the fields leave the reconstructed curve, we use}
\begin{equation}
 {V(\phi,\psi)=U(\phi)+J(\phi)[\psi-\psi_{\rm tr}(\phi)]+\frac12\kappa_\perp[\psi-\psi_{\rm tr}(\phi)]^2,}
\label{eq:potential_completion}
\end{equation}
 {Here $\psi=\psi_{\rm tr}(\phi)$ is the reconstructed trajectory; along its monotonic branch, $U(\phi)\equiv V[z(\phi)]$ and $\psi_{\rm tr}(\phi)\equiv\psi[z(\phi)]$.  The linear term specifies the force in the $\psi$ direction at fixed $\phi$; the quadratic term specifies how this force changes away from the curve.  Choosing $J=[\ddot\psi+3H\dot\psi]_{t=t(\phi)}$ gives, on the trajectory,
\begin{equation}
 V_{,\psi}=J,\qquad
 V_{,\phi}=U_{,\phi}-J\psi_{{\rm tr},\phi}
 =\frac{\dot V-J\dot\psi}{\dot\phi}.
\label{eq:potential_gradient_on_path}
\end{equation}
The first relation enforces the phantom Klein--Gordon equation.  Since $S_K>|K|$ keeps $\dot\phi\ne0$, the dark-energy continuity equation and the second relation give the canonical equation.  Equation~(\ref{eq:potential_completion}) is a fixed function of the field positions; redshift labels only the path used to reconstruct $U$, $J$, and $\psi_{\rm tr}$.  For the reference realization we set $\varsigma=0.50$ and $\kappa_\perp=-\mu_\perp^2$ with $\mu_\perp/H_0=2$, which gives the phantom equation a restoring contribution for a displacement in $\psi$ at fixed $\phi$.  Here transverse means varying $y=\psi-\psi_{\rm tr}(\phi)$ at fixed $\phi$, not an orthogonal projection with the indefinite field-space metric.  The coefficient $\mu_\perp$ is a scalar mass scale, not a mass eigenvalue of the coupled system.  Changing this curvature leaves the value and gradient of $V$ on $y=0$ unchanged.  Thus $\varsigma$ and $\mu_\perp$ specify additional choices in the field realization; the MEDE chains do not constrain them.}

{
For the reconstructed functions, the potential is nonseparable: the force on either field depends on the other field's position.  For constant $\kappa_\perp$, its mixed derivative is
\begin{equation}
 V_{,\phi\psi}=J_{,\phi}-\kappa_\perp\psi_{{\rm tr},\phi}
 =J_{,\phi}+\mu_\perp^2\psi_{{\rm tr},\phi},
 \label{eq:mixed_potential}
\end{equation}
where the second equality uses $\kappa_\perp=-\mu_\perp^2$.  The potential gradients therefore couple the two rolling fields.  In particular, phantom-sign kinetic dominance does not imply domination by a separately conserved phantom fluid: the two fields share the interaction potential in Eq.~(\ref{eq:potential_completion}).  Differentiating $K=(\dot\phi^2-\dot\psi^2)/2$ and using their Klein--Gordon equations gives
\begin{equation}
 \dot K+6HK=-\dot V,\qquad
 \dot V=V_{,\phi}\dot\phi+V_{,\psi}\dot\psi.
 \label{eq:kinetic_balance}
\end{equation}
For $\rho_q>0$, the sign of $K$ fixes the side of the phantom divide.  At $K=0$, the crossing direction is set by the change in the shared potential energy.  On the reconstructed MEDE trajectory, $\dot K|_{z_t}=\Delta H(z_t)\rho_{\rm de}(z_t)/2>0$ for $\Delta>0$, so $\dot V|_{z_t}<0$ and the crossing proceeds from phantom to nonphantom evolution.
}

The interaction in Eq.~(\ref{eq:mixed_potential}) is a force coupling, with no phenomenological decay rate.  Both fields remain present during the crossing.  To define separate equations of state, one would first have to assign the shared potential energy to the two fields.  For example, an arbitrary smooth allocation function $F(\phi,\psi)$ gives
\begin{align*}
&\rho_\phi=\tfrac12\dot\phi^2+F,\qquad
p_\phi=\tfrac12\dot\phi^2-F,\\
&\rho_\psi=-\tfrac12\dot\psi^2+V-F,\qquad
p_\psi=-\tfrac12\dot\psi^2-V+F .
\end{align*}
Then $w_\phi=p_\phi/\rho_\phi$ and $w_\psi=p_\psi/\rho_\psi$ depend on the allocation wherever their denominators are nonzero.  An energy-transfer rate inferred from their separate continuity equations would also depend on $F$.  Changing $F$ leaves $\rho_q$, $p_q$, and the field equations unchanged.  We therefore describe the crossing through the total $w_q$ and the kinetic-energy difference, and show the work done by the potential on each field in Fig.~\ref{fig:forward_dynamics}(d).  These quantities specify the energy balance independently of the allocation $F$.

\paragraph{Potential-energy release at the crossing.}
Energy conservation gives $\rho_{q,N}=-3(1+w_q)\rho_q=-6K$, where $N=\ln a$ and the notation ",N" denotes differentiation with respect to $N$.  At $w_q=-1$, differentiating this relation and using Eq.~(\ref{eq:kinetic_balance}) gives
\begin{equation}
\left.-\frac{\dot V}{H\rho_q}\right|_{w_q=-1}
=\left.\frac12\frac{\dd w_q}{\dd N}\right|_{w_q=-1}
=\left.-\frac{\rho_{q,NN}}{6\rho_q}\right|_{w_q=-1}.
\label{eq:turnover_work}
\end{equation}
This dimensionless rate, $\mathcal R_t$, is positive at a simple density maximum and independent of the choices of $S_K$ and $\kappa_\perp$.  For the natural-logarithm MEDE history it equals $\Delta/2$, with posterior values in Table~\ref{tab:full_results}.  The identity fixes the net work along the path; choosing the kinetic split determines the separate forces.  Suitable separable quintom potentials can also give the same crossing direction~\cite{2025MNRAS.544.3142G,2026MNRAS.551g1403G}.

\paragraph{Field-space geometry and local reconstruction freedom.}
The fields take values on a two-dimensional manifold $\mathcal M$ with the fixed, indefinite metric $G_{IJ}=\mathrm{diag}(1,-1)$.  Their homogeneous evolution defines a curve $\gamma:t\mapsto(\phi(t),\psi(t))$.  The expansion determines the pullbacks of the metric and potential to that curve,
\begin{equation}
\begin{aligned}
&\gamma^*G=(\rho_q+p_q)\,\dd t^2,\\
&\gamma^*V=V[\gamma(t)]=\frac{\rho_q-p_q}{2}.
\end{aligned}
\label{eq:field_pullbacks}
\end{equation}
Here a pullback is the metric or potential evaluated along the field trajectory. Equation~(\ref{eq:field_pullbacks}) determines the kinetic-energy difference and the potential on the curve, while the individual velocities and the curve itself remain free.  At the crossing, the tangent $\dot\gamma$ is nonzero but null: $G_{IJ}\dot q^I\dot q^J=0$.  The field-space metric remains nondegenerate.  An orthonormal tangent--normal frame based on division by $\sqrt{|2K|}$ is singular there, so we use the regular coordinates $(\phi,y)$ with $y=\psi-\psi_{\rm tr}(\phi)$.  Their Jacobian with respect to $(\phi,\psi)$ is unity.  Thus $y$ measures departure from the curve at fixed $\phi$, not metric-orthogonal distance.

Fix a regular, non-self-intersecting segment of this trajectory.  If two smooth potentials produce the same homogeneous fields and expansion with $G_{IJ}$ and the standard species fixed, their difference $\mathcal F\equiv\widetilde V-V$ and its first derivatives vanish at $y=0$.  Repeated application of Hadamard's lemma, or Taylor's formula with integral remainder in $y$, gives
\begin{equation}
\mathcal F(\phi,y)=y^2\int_0^1(1-\xi)\mathcal F_{,yy}(\phi,\xi y)\,\dd\xi
=\tfrac12 B(\phi)y^2+y^3C(\phi,y),
\label{eq:local_potential_family}
\end{equation}
where $B(\phi)=\mathcal F_{,yy}(\phi,0)$ and $C$ is smooth.  The vanishing value of $\mathcal F$ and its first derivative give the factor $y^2$; separating its coefficient at $y=0$ leaves the cubic remainder.  If the Hessians also agree, $B=0$.  This is the general local form for the fixed trajectory, with no assumption about the potential outside its neighborhood.  The shared potential value, gradient, and Hessian comprise its second-order jet along $\gamma$; matching them is sufficient to preserve the background and linear field equations.

For two or more fields with a smooth fixed nondegenerate metric, analogous local freedom exists in coordinates transverse to a regular trajectory.  This statement concerns reconstruction, not the existence of a crossing for every metric.  Positive-definite $G_{IJ}$ and positive field density imply $\rho_q+p_q\geq0$ for minimally coupled fields.  Our crossing construction uses the indefinite metric specified above.

For the potential deformation in Eq.~(\ref{eq:potential_freedom}), the added value and gradient vanish at $y=0$.  The same fields, initial data, and standard-sector content therefore remain an exact background solution.  Defining $s(\phi)\equiv\dd\psi_{\rm tr}/\dd\phi$, the Hessian changes on this path by
\begin{equation}
 \left.(\widetilde V_{,IJ}-V_{,IJ})\right|_{y=0}
 =B(\phi)\begin{pmatrix}s^2&-s\\-s&1\end{pmatrix},
 \label{eq:curvature_freedom}
\end{equation}
with rows and columns ordered as $(\phi,\psi)$.  Equation~(\ref{eq:curvature_freedom}) acts only on the departure $\delta y=\delta\psi-s\delta\phi$ from the field path.  A common time shift gives $\delta y=0$, since $\dot\psi=s\dot\phi$; the definition remains regular at the crossing.  If $B=0$, the Hessian is unchanged too.  The reference background and all linear perturbations, for the same initial perturbations, are then identical even though the potentials differ away from the path.  Finite displacements and nonlinear evolution can distinguish them.  Away from the reference trajectory, the dynamics and stability depend on $B$, $C$, and the field initial state.

\paragraph{An explicit family with different off-trajectory forces.}
Figure~\ref{fig:equivalent_potentials} uses the local family
\begin{equation}
V_\lambda(\phi,\psi)=V(\phi,\psi)+\lambda\rho_{{\rm de},0}\left(\frac{y}{L}\right)^3,
\qquad L=0.1M_{\rm Pl},\qquad \lambda=0,\pm0.1.
\label{eq:cubic_example}
\end{equation}
The scale $L$ and dimensionless amplitudes are chosen to illustrate the freedom, with all other functions and parameters held fixed.  For $s=\dd\psi_{\rm tr}/\dd\phi$, the changes in the two potential-force terms of the Klein--Gordon equations are
\begin{equation}
-\bigl(V_{\lambda,\phi}-V_{,\phi}\bigr)
=\frac{3\lambda\rho_{{\rm de},0}}{L^3}s\,y^2,
\qquad
V_{\lambda,\psi}-V_{,\psi}
=\frac{3\lambda\rho_{{\rm de},0}}{L^3}y^2.
\label{eq:cubic_force}
\end{equation}
Both force differences start at second order in the displacement.  The added third derivatives on the path are $6\lambda\rho_{{\rm de},0}n_In_Jn_K/L^3$, where $n_I=(-s,1)$.  Thus these potentials share the same homogeneous solution and linear fluctuations but have different higher-order field interactions.  The panels evaluate the potential and its phantom-direction derivative at fixed $\phi=\phi(z_t)$ over $|y|\leq0.05M_{\rm Pl}$.  They do not represent separate nonlinear cosmological evolutions or constrain $\lambda$.  Equality of the potential, gradient, and Hessian on the entire path follows analytically from Eq.~(\ref{eq:cubic_example}); direct evaluation of the archived potential coefficients verifies it.  This is a field-space example of the nonlinear reconstruction freedom established in scalar--tensor theories~\cite{2017PhRvD..96h4051K,2019PhRvD.100d4034K}, here with the kinetic terms fixed and a phantom-divide crossing.

The reconstruction applies to any smooth prescribed history with $\rho_{\rm de}>0$ on the chosen interval.  Its input is $K=(1+w)\rho_{\rm de}/2$ and $V=(1-w)\rho_{\rm de}/2$, followed by the kinetic split and the local potential extension.  These steps do not depend on the functional form used to describe the history.

Figure~\ref{fig:quintom_reconstruction} gives the reconstructed kinetic-energy difference, potential energy, and field profiles for the reference history.

\begin{figure*}[t]
\centering
\includegraphics[width=0.88\textwidth]{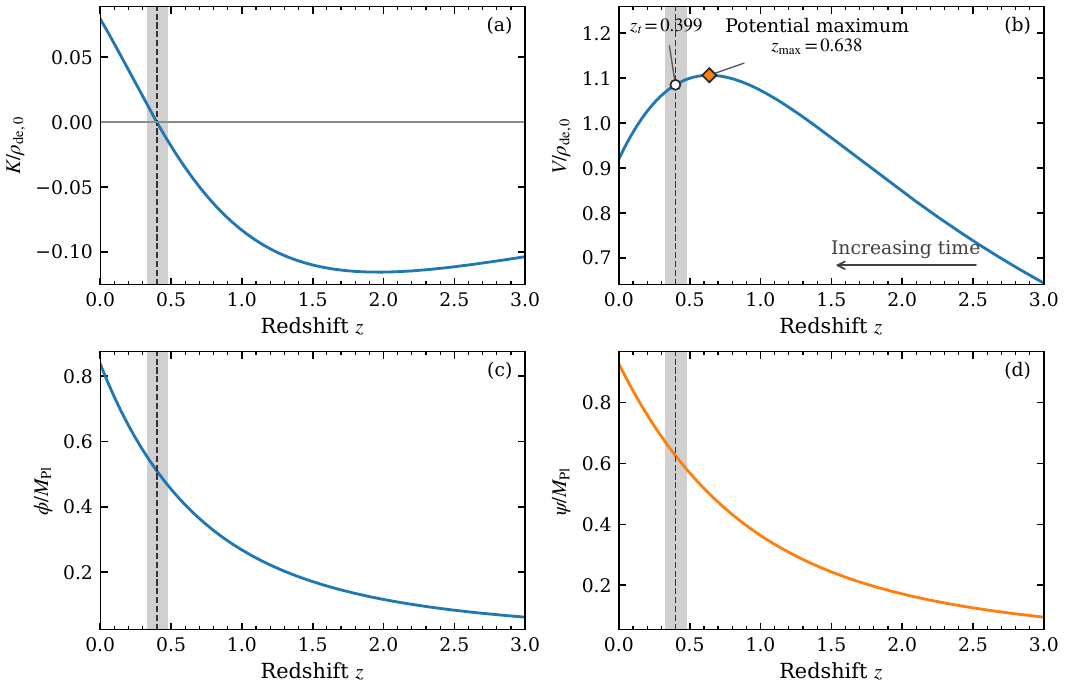}
\caption{The scalar-quintom reconstruction uses the representative realization $z_t=0.399$, $\Delta=0.492$, and $\Omega_{\rm m0}=0.3121$, corresponding to the joint CMB+DESI DR2 BAO+DES-Dovekie posterior median.  Upper panels show (a) the kinetic-energy difference $K(z)$ and (b) the potential $V(z)$, normalized by $\rho_{{\rm de},0}$; lower panels show (c) $\phi$ and (d) $\psi$ in reduced Planck-mass units.  In (b), the diamond marks the potential maximum at $z_{\max}=0.638$ and the open circle marks the crossing at $z_t=0.399$, with the same symbols as Fig.~\ref{fig:forward_dynamics}(a).  Increasing cosmic time runs from right to left.  Dashed lines mark $z_t$; gray bands show its marginalized $68\%$ interval, not uncertainty in the field profiles.  The profiles use $\varsigma=0.50$ and the field origin $\phi=\psi=0$ at $N=-10$; this coordinate shift has no observable significance.}
\label{fig:quintom_reconstruction}
\end{figure*}

\begin{figure*}[t]
\centering
\includegraphics[width=0.49\textwidth]{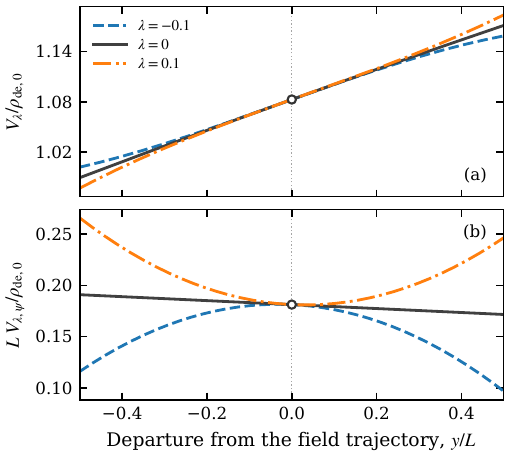}
\caption{Different local interactions with the same background and linear perturbations.  (a) $V_\lambda=V+\lambda\rho_{{\rm de},0}(y/L)^3$ and (b) its potential force in the phantom equation, at fixed $\phi=\phi(z_t)$ for the reference solution, with $L=0.1M_{\rm Pl}$.  Open circles mark the common trajectory, $y=0$.  The added value, gradient, and Hessian vanish there at every $\phi$, so the equality of the background and linear equations is exact.  The illustrative choices $\lambda=0,\pm0.1$ show how finite field displacements distinguish the potentials; they are not fitted parameters.}
\label{fig:equivalent_potentials}
\end{figure*}

\suppsection{Forward field evolution}
\label{app:forward}

The functions $U(\phi)$, $J(\phi)$, and $\psi_{\rm tr}(\phi)$ specify the potential in Eq.~(\ref{eq:potential_completion}).  With this potential fixed, the field positions and velocities obey
\begin{equation}
 \ddot q^I+3H\dot q^I+G^{IJ}V_{,J}=0,
\label{eq:forward_kg}
\end{equation}
where $q^I=(\phi,\psi)$ and $G_{IJ}={\rm diag}(1,-1)$.  Explicitly, the equations are $\ddot\phi+3H\dot\phi+V_{,\phi}=0$ and $\ddot\psi+3H\dot\psi-V_{,\psi}=0$.  The Friedmann equations are
\begin{equation}
3\Mpl^2H^2=\rho_{\rm std}+\rho_q,\qquad
\dot H=-\frac{\rho_{\rm std}+p_{\rm std}+\dot\phi^2-\dot\psi^2}{2\Mpl^2}.
\label{eq:forward_friedmann}
\end{equation}
The standard species include baryons, cold dark matter, photons, massless relics, and the massive neutrino with its time-dependent density and pressure.  The field equation of state is calculated from the evolving fields,
\begin{equation}
w_q=\frac{\dot\phi^2/2-\dot\psi^2/2-V}{\dot\phi^2/2-\dot\psi^2/2+V}.
\label{eq:field_equation_of_state}
\end{equation}
The same quantity can be expressed in terms of the expansion and the specified standard components:
\begin{align}
 &\rho_H(z)=3M_{\rm Pl}^2H_0^2E^2(z)-\rho_{\rm std}(z),\label{eq:rho_from_H}\\
 &w_H(z)=\frac{3M_{\rm Pl}^2H_0^2[\frac23(1+z)EE_{,z}-E^2]-p_{\rm std}(z)}
 {3M_{\rm Pl}^2H_0^2E^2-\rho_{\rm std}(z)}.\label{eq:w_from_H}
\end{align}
Equations~(\ref{eq:rho_from_H}) and (\ref{eq:w_from_H}) make the matter-density dependence explicit.  The same $H(z)$ can imply different dark-energy equations of state if different matter densities are subtracted.

We integrate with $N=\ln a$ as the independent variable and represent the reconstructed functions by cubic splines.  For any twice-differentiable background quantity $X(N)$, cosmic-time derivatives are obtained from
\begin{equation}
\dot X=HX_{,N},\qquad
\ddot X=H^2\left[X_{,NN}+\frac{H_{,N}}{H}X_{,N}\right].
\label{eq:time_conversion}
\end{equation}
The representative realization has $z_t=0.399$, $\Delta=0.492$, and $\Omega_{\rm m0}=0.3121$. We use $\varsigma=0.50$ and $\mu_\perp/H_0=2$.

\paragraph{Initial data and integration.}
\label{app:native_class}
The background field equations are solved using the CLASS code~\cite{2011JCAP...07..034B}, evaluating the potential and its derivatives at the current field positions.  The resulting background and Hessian $V_{,IJ}$ supply the coefficients of the perturbation equations in Sec.~\ref{app:perturbations}.  Photons and neutrinos evolve with their Boltzmann hierarchies.

For the reference integration, we extend the reconstructed functions into radiation domination and choose field origins such that $\phi,\psi\to0$ as $a\to0$.  A change of field origin translates the potential and initial positions together, leaving the dynamics unchanged.  At $a_i=10^{-14}$ the initial data, quoted to three significant figures, are
\begin{equation}
\phi_i/\Mpl=2.69\times10^{-36},\qquad
\psi_i/\Mpl=4.15\times10^{-36},\qquad
\phi_{,N,i}/\phi_i=\psi_{,N,i}/\psi_i=2.67 .
\label{eq:native_initial_data}
\end{equation}
These values select the reconstructed rolling solution.  The reference cosmology used in the CLASS code has $H_0=67.5\,{\rm km\,s^{-1}\,Mpc^{-1}}$, $\omega_b=0.0222$, $\omega_{\rm cdm}=0.119$, one $0.06\,{\rm eV}$ massive neutrino, and $N_{\rm eff}\simeq3.046$.  The primordial and reionization parameters are $\ln(10^{10}A_s)=3.05$, $n_s=0.967$, and $\tau_{\rm reio}=0.0563$.  Here $\omega_i=\Omega_{i0}h^2$ and $h=H_0/(100\,{\rm km\,s^{-1}\,Mpc^{-1}})$. Over $0\leq z\leq3$, the maximum fractional difference in $H$ and the maximum absolute difference in $w_q$ from the target MEDE solution are both of order $10^{-8}$.
 
\paragraph{Early-time continuation.}
The radiation-era potential follows from extending the chosen MEDE history to high redshift.  This extension is a model assumption: low-redshift distances do not directly reconstruct dark energy during radiation domination.  With $\rho_q\propto a^{3\Delta}$, $H\propto a^{-2}$, and the fixed kinetic split, the rolling branch satisfies $\dot\phi,\dot\psi\propto a^{3\Delta/2}$ and $\phi,\psi\propto a^{2+3\Delta/2}$.  Its gradients scale as $|V_{,I}|\propto a^{-2+3\Delta/2}$, while $|V_{,I}|/(H^2M_{\rm Pl})\propto a^{2+3\Delta/2}$ tends to zero.  Large dimensionful gradients can therefore coexist with negligible field density and strongly Hubble-damped motion.

Along the trajectory, $U(\phi)\propto\phi^{6\Delta/(4+3\Delta)}$, approximately $\phi^{0.539}$ for $\Delta=0.492$.  The slope diverges at the formal zero-field endpoint.  The calculation uses a finite tabulated domain and does not extend through that endpoint.  A different early-time continuation would define a different potential and would require its own evolution.

\suppsection{Linear field perturbations}
\label{app:perturbations}

The two field perturbations are evolved with the matter, radiation, and metric perturbations in the synchronous gauge used by the CLASS code.  The metric is
\begin{equation}
 \dd s^2=a^2(\tau)[-\dd\tau^2+(\delta_{ij}+h_{ij})\dd x^i\dd x^j],\qquad
 h_{ij}=\hat k_i\hat k_jh_{\rm s}+6\left(\hat k_i\hat k_j-\frac13\delta_{ij}\right)\eta ,
\label{eq:synchronous_metric}
\end{equation}
where $\tau$ is conformal time, $\delta_{ij}$ is the Euclidean spatial metric, and $\hat k_i$ is the unit vector along the Fourier wavevector.  The scalar metric modes are $h_{\rm s}$ and $\eta$.  A prime denotes $\dd/\dd\tau$, and $\mathcal H=a'/a=aH$.  The field perturbations satisfy
\begin{align}
 &\delta\phi''+2{\cal H}\delta\phi'+(k^2+a^2V_{,\phi\phi})\delta\phi
 +a^2V_{,\phi\psi}\delta\psi=-\tfrac12\phi'h_{\rm s}',
\label{eq:synchronous_phi}\\
 &\delta\psi''+2{\cal H}\delta\psi'+(k^2-a^2V_{,\psi\psi})\delta\psi
 -a^2V_{,\phi\psi}\delta\phi=-\tfrac12\psi'h_{\rm s}'.
\label{eq:synchronous_psi}
\end{align}
Here $k$ is the comoving wavenumber.  The metric drives perturbations in the rolling fields, and the mixed derivative of the potential couples their fluctuations.  The potential curvature enters with opposite signs because the two kinetic terms have opposite signs.

The density, pressure, and momentum perturbations are
\begin{align}
 &\delta\rho_q=a^{-2}(\phi'\delta\phi'-\psi'\delta\psi')
 +V_{,\phi}\delta\phi+V_{,\psi}\delta\psi,
\label{eq:field_density_perturbation}\\
 &\delta p_q=a^{-2}(\phi'\delta\phi'-\psi'\delta\psi')
 -V_{,\phi}\delta\phi-V_{,\psi}\delta\psi,
\label{eq:field_pressure_perturbation}\\
 &(\rho_q+p_q)\theta_q=\frac{k^2}{a^2}(\phi'\delta\phi-\psi'\delta\psi),\qquad \sigma_q=0.
\label{eq:field_momentum_perturbation}
\end{align}
The convention is $(\rho_q+p_q)\theta_q\equiv ik^j\delta T^0{}_j$, where $\theta_q$ denotes the scalar velocity divergence of the field component, distinct from the background expansion scalar. $\sigma_q$ denotes the scalar anisotropic stress. The Einstein equations use the momentum density directly.  None of Eqs.~(\ref{eq:synchronous_phi})--(\ref{eq:field_momentum_perturbation}) divides by $\rho_q+p_q$, so these variables remain regular at the crossing~\cite{2005PhRvD..72l3515Z,2010PhR...493....1C}.

The standard cosmic species start in the regular adiabatic mode.  We set the independent initial field fluctuations and their derivatives to zero when the field density is negligible.  This specifies the absence of a primordial field-isocurvature mode; the later field response is sourced by the evolving metric.  Primordial field fluctuations would supply additional initial data, not fixed by the reconstructed background.

The potential enters the background through its value and gradient, and the linear equations through these quantities and the Hessian.  This explains the distinction between quadratic and cubic additions in Eq.~(\ref{eq:potential_freedom}).  A nonzero $B$ changes the Hessian by Eq.~(\ref{eq:curvature_freedom}), while $y^3C(\phi,y)$ and its first two derivatives vanish on the reference trajectory.  For identical initial perturbations, this cubic freedom leaves the field, metric, and standard-species perturbations unchanged.

\suppsection{Linear growth observables}
\label{app:observables}

\suppsubsection{DESI ShapeFit growth prediction}
\label{app:shapefit}

The anisotropy of galaxy clustering measures the velocity response to the growing matter fluctuations.  DESI DR1 ShapeFit reports a compressed growth coordinate $f\sigma_{s8}$~\cite{2025JCAP...09..008A}.  For Fig.~\ref{fig:field_linear_shapefit}, we estimate it from the growth rate and linear power at the template pivot:
\begin{equation}
d_f(z)\equiv\frac{f\sigma_{s8}}{(f\sigma_{s8})_{\rm fid}}
=\frac{f\sqrt{A_p}}{[f\sqrt{A_p}]_{\rm fid}},
\qquad
A_p=s^{-3}P_{cb}(k_p/s,z),\qquad
s=\frac{r_dh}{(r_dh)_{\rm fid}} .
\label{eq:shapefit_growth}
\end{equation}
Here $P_{cb}$ is the cold-dark-matter-plus-baryon spectrum, $k_p=0.03\,h\,{\rm Mpc}^{-1}$ is the template pivot, and $\mathrm{fid}$ denotes the DESI fiducial template.  For this comparison we estimate $f$ as $\tfrac12\partial\ln P_{cb}/\partial\ln a$ at the dilated pivot. The DESI likelihood instead estimates $f$ from the velocity and density spectra.  The subscript $s8$ labels the ShapeFit normalization; this compressed quantity is not identified with the ordinary $f\sigma_8$ measured at a fixed $8h^{-1}\,{\rm Mpc}$ smoothing radius.

We hold the cosmological parameters and the primordial spectrum fixed in the three realizations in Fig.~\ref{fig:field_linear_shapefit}. Increasing $\mu_\perp/H_0$ from $2$ to $4$ at $\varsigma=0.50$ changes the potential Hessian but leaves the reference path and initial field state unchanged. Increasing $\varsigma$ from $0.50$ to $1.00$ at $\mu_\perp/H_0=2$ instead changes how the two fields share the kinetic terms. In this case we reconstruct the path and potential and use the corresponding initial field values and velocities. The forward solutions recover the same expansion history. The spread in $f\sigma_{s8}$ at the six DESI redshifts is much smaller than the marginal measurement errors. Table~\ref{tab:shapefit_growth} lists the measurements and the reference prediction.

To draw the curves, we evaluate $P_{cb}$ on a grid in redshift and differentiate its logarithm at the pivot to obtain $f$. We interpolate the ratio of the published fiducial $f\sigma_{s8}$ to the computed fiducial $f\sqrt{A_p}$ between the six effective redshifts and hold that ratio fixed at the endpoint values outside this interval. The ratio varies by less than $0.06\%$ over the plotted range, so the turnover is set by the product $f\sqrt{A_p}$. The reference curve recovers the six predictions to the precision shown in Table~\ref{tab:shapefit_growth}.

\begin{table}[!htbp]
\centering
\caption{DESI DR1 ShapeFit growth measurements and the direct field prediction for the reference realization. Measurement errors are marginal $1\sigma$ errors.}
\label{tab:shapefit_growth}
\begin{tabular}{lccc}
\toprule
Tracer & $z_{\rm eff}$ & Measured $f\sigma_{s8}$ & Field prediction\\
\midrule
BGS & 0.295 & $0.396\pm0.090$ & 0.460 \\
LRG1 & 0.510 & $0.548\pm0.059$ & 0.465 \\
LRG2 & 0.706 & $0.480\pm0.048$ & 0.458 \\
LRG3 & 0.919 & $0.438\pm0.041$ & 0.441 \\
ELG2 & 1.317 & $0.373\pm0.034$ & 0.399 \\
QSO & 1.491 & $0.437\pm0.046$ & 0.380 \\
\bottomrule
\end{tabular}
\end{table}

For the reference realization, the largest difference between the field prediction and a measured value is $1.38$ times its marginal error. This is a comparison of individual residuals, not a joint goodness-of-fit test.

\suppsection{Physical interpretation}
\label{app:scope}

The reconstructed potential is defined near a chosen field trajectory.  The kinetic split and the extension away from that trajectory specify a classical realization of the inferred history.  Its crossing occurs in the actual field stress-energy, $w_q=p_q/\rho_q$, with cold dark matter separately conserved.  This differs from an apparent crossing inferred after assigning an incorrect matter contribution to the expansion~\cite{2026JCAP...05..008C,2026arXiv260716191K,2026ApJ..1010L..12J}.

The distinction matters when comparing explanations of the inferred history.  Interacting dark-energy models exchange energy with dark matter~\cite{2024PhRvL.133y1003G,2026arXiv260107361L,2024ApJ...976....1L,2026PhRvD.113b3515P,2025JCAP...12..018L,2025SCPMA..6910413L}, whereas the interaction reconstructed here acts within the two-field sector.  Nonminimal and scalar-tensor couplings alter the relation between expansion, field stress-energy, and the metric response~\cite{2025PhRvL.134r1002Y,2026PhRvD.113d1301T,2025PhRvL.135h1001W,2026PhRvD.114d3514W}; derivative interactions provide further possibilities~\cite{2026PhRvD.113l2006C,2024PhRvD.110l3524C}.  These alternatives can share a similar expansion while differing in their perturbations and stability.  References~\cite{2026RAA....26h4002L,Zhang:2026zbs} review the broader range of interpretations.

The negative kinetic energy of $\psi$ makes the Hamiltonian unbounded below.  Quantizing this degree of freedom permits vacuum decay into negative-energy field excitations and ordinary particles~\cite{2010PhR...493....1C}.  Regular classical evolution through $w_q=-1$ does not remove this problem.  The present construction is consequently restricted to classical dynamics on the specified domain.  A viable quantum theory requires additional physical structure beyond the reconstructed potential.

Effective theories of dark energy organize the kinetic and gradient terms needed to assess such alternatives~\cite{2014JCAP...07..050B,2020PhR...857....1F}.  A candidate theory must avoid ghost and gradient instabilities, and its tensor propagation must satisfy the multimessenger constraint from GW170817~\cite{2017PhRvL.119y1302C}.  These properties depend on terms in the action that the reconstructed background does not determine.

Observations of the expansion constrain the total dark-energy density and pressure under the assumed matter content and gravity theory.  Linear clustering also depends on the potential curvature and the initial field perturbations.  The cubic freedom in Eq.~(\ref{eq:potential_freedom}) survives both levels of information.  Resolving it requires information sensitive to higher derivatives of the potential, such as nonlinear field evolution, or additional assumptions about the interaction.

\end{document}